\RequirePackage{ifpdf}
\ifpdf 
\documentclass[pdftex]{sigma}
\else
\documentclass{sigma}
\fi

\newcommand{\comp}{\leftrightarrow}

\newcommand{\ord}{\operatorname{ord}}
\def\comp{\leftrightarrow}
\newcommand{\notcomp}{\not\kern-0.2ex\comp}
\newcommand{\oo}{\stackrel{(o)}{\longrightarrow}}
\def\too{\stackrel{(o)}{\longrightarrow}}

\begin{document}

\allowdisplaybreaks

\renewcommand{\thefootnote}{$\star$}

\renewcommand{\PaperNumber}{003}

\FirstPageHeading

\ShortArticleName{Modularity, Atomicity and States in Archimedean Lattice Ef\/fect Algebras}

\ArticleName{Modularity, Atomicity and States \\ in Archimedean Lattice Ef\/fect Algebras\footnote{This paper is a
contribution to the Proceedings of the 5-th Microconference
``Analytic and Algebraic Me\-thods~V''. The full collection is
available at
\href{http://www.emis.de/journals/SIGMA/Prague2009.html}{http://www.emis.de/journals/SIGMA/Prague2009.html}}}

\Author{Jan PASEKA}

\AuthorNameForHeading{J. Paseka}

\Address{Department of Mathematics
 and Statistics,
Faculty of Science,
Masaryk University,\\
Kotl\'a\v{r}sk\'a~2,
CZ-611~37~Brno, Czech Republic}
\Email{\href{mailto:paseka@math.muni.cz}{paseka@math.muni.cz}}

\ArticleDates{Received September 29, 2009, in f\/inal form January 07, 2010;  Published online January 08, 2010}

\Abstract{Ef\/fect algebras are a generalization of many structures
which arise in quantum physics and in  mathematical
economics. We show that, in every
modular Archime\-dean atomic lattice ef\/fect algebra $E$ that is not
an orthomodular lattice there exists an $(o)$-continuous state $\omega$ on $E$,
which is subadditive.  Moreover, we show properties of f\/inite and
compact elements of such lattice ef\/fect algebras.}


\Classification{06C15;  03G12; 81P10}

\section{Introduction, basic def\/initions and some known facts}\label{intro}

Ef\/fect algebras (introduced by D.J.~Foulis and M.K.~Bennett in
\cite{FoBe} for modelling unsharp measurements in a Hilbert space)
may be carriers of states or probabilities when events
are noncompatible or unsharp resp. fuzzy. In this setting, the
set $\cal E(H)$ of ef\/fects on a Hilbert space $\cal H$ is the set
of all Hermitian operators on $\cal H$ between the
null operator~$0$ and the identity
operator~$1$, and the partial operation $\oplus$ is the
restriction of the usual operator sum.
D.J.~Foulis and M.K.~Bennett recognized that
ef\/fect algebras are equivalent to D-posets introduced
in general form by F.~K\^opka and  F.~Chovanec
(see~\cite{KoCh}), f\/irstly def\/ined as axiomatic
systems of fuzzy sets by F.~K\^opka in~\cite{Kop}.

Ef\/fect algebras are a generalization of many structures
which arise in quantum physics (see~\cite{BGL}) and in  mathematical
economics  (see~\cite{dvurec2,EpZh}). There are some
basic ingredients in the study of the mathematical
foundations of physics, typically  the
fundamental concepts are states, observables and symmetries.
These concepts are tied together in \cite{foulis2007} by employing ef\/fect
algebras.

It is a remarkable fact that there are
even f\/inite ef\/fect algebras admitting no states, hence no probabilities.
The smallest of them has only nine elements (see~\cite{ZR62}).  One
possibility for eliminating this unfavourable situation is to
consider modular complete lattice ef\/fect algebras
(see~\cite{ZR74}).

Having this in mind, we are going to show that, in every
modular Archimedean atomic lattice ef\/fect algebra~$E$ that is not
an orthomodular lattice there exists an $(o)$-continuous state~$\omega$ on~$E$,
which is subadditive.

We show some further important properties
of f\/inite elements in modular lattice ef\/fect al\-geb\-ras. Namely, the
set $G$ of all f\/inite elements
in a modular lattice ef\/fect algebra  is a lattice ideal of $E$.
Moreover, any compact element in an Archimedean lattice ef\/fect algebra
$E$  is a f\/inite join of f\/inite elements of $E$.

\begin{definition}[\cite{FoBe}]\label{def:EA}
A partial algebra $(E;\oplus,0,1)$ is called an {\em effect algebra} if
$0$, $1$ are two distinct elements and $\oplus$ is a partially
def\/ined binary operation on $E$ which satisfy the following
conditions for any $x,y,z\in E$:
\begin{description}\itemsep=0pt
\item[$(Ei)$\phantom{ii}] $x\oplus y=y\oplus x$ if $x\oplus y$ is def\/ined,
\item[$(Eii)$\phantom{i}] $(x\oplus y)\oplus z=x\oplus(y\oplus z)$  if one
side is def\/ined,
\item[$(Eiii)$] for every $x\in E$ there exists a unique $y\in
E$ such that $x\oplus y=1$ (we put $x'=y$),
\item[$(Eiv)$\phantom{i}] if $1\oplus x$ is def\/ined then $x=0$.
\end{description}
\end{definition}

We put $\bot =\{(x, y)\in E\times E \mid x\oplus y\ \mbox{is def\/ined}\}$.
We often denote the ef\/fect algebra $(E;\oplus,0,1)$ brief\/ly by
$E$. On every ef\/fect algebra $E$  the partial order
$\le$  and a partial binary operation $\ominus$ can be
introduced as follows:
\[
x\le y \mbox{ and }  y\ominus x=z  \mbox{ if\/f } x\oplus z
\mbox{ is def\/ined and } x\oplus z=y .
\]

Elements $x$ and $y$ of an ef\/fect algebra $E$   are said to be
{\em $($Mackey$)$ compatible} ($x\leftrightarrow y$ for short) if\/f there
exist elements $x_1, y_1, d \in E$ with $x = x_1 \oplus d$, $y = y_1 \oplus d$ and
$x_1 \oplus y_1 \oplus d\in E$.

If $E$ with the def\/ined partial order is a lattice (a complete
lattice) then $(E;\oplus,0,1)$ is called a {\em lattice effect
algebra} ({\em a complete lattice effect algebra}).

If, moreover, $E$ is a modular or distributive lattice
then $E$ is called {\em modular} or {\em distributive} ef\/fect
algebra.

Lattice ef\/fect algebras generalize two important structures:
orthomodular lattices and $MV$-algebras. In fact a lattice ef\/fect
algebra $(E;\oplus,0,1)$ is an orthomodular lattice \cite{kalmbach} if\/f
$x\wedge x'=0$ for every $x\in E$ (i.e., every $x\in E$ is a {\em sharp
element\/}). A lattice ef\/fect algebra can be organized into an
$MV$-algebra \cite{C.C.Ch} (by extending $\oplus$ to a total binary operation
on $E$) if\/f  any two elements of $E$ are compatible if\/f
$(x\vee y)\ominus x=y\ominus(x\wedge y))$  for every pair
of elements $x, y\in E$ \cite{Kop2,Kop3}.

A minimal nonzero element of an ef\/fect algebra  $E$
is called an {\em atom}  and $E$ is
called {\em atomic} if under every nonzero element of
$E$ there is an atom.

We say that a f\/inite system $F=(x_k)_{k=1}^n$ of not necessarily
dif\/ferent elements of an ef\/fect algebra $(E;\oplus,0,1)$ is
$\oplus $-{\it orthogonal} if $x_1\oplus x_2\oplus \cdots\oplus
x_n$ (written $\bigoplus\limits_{k=1}^n x_k$ or $\bigoplus F$) exists
in $E$. Here we def\/ine $x_1\oplus x_2\oplus \cdots\oplus x_n=
(x_1\oplus x_2\oplus \cdots\oplus x_{n-1})\oplus x_n$ supposing
that $\bigoplus\limits_{k=1}^{n-1}x_k$ is def\/ined and
$\bigoplus\limits_{k=1}^{n-1}x_k\le x'_n$. We also def\/ine
$\bigoplus \varnothing=0$.
An arbitrary system
$G=(x_{\kappa})_{\kappa\in H}$ of not necessarily dif\/ferent
elements of $E$ is called $\oplus $-orthogonal if $\bigoplus K$
exists for every f\/inite $K\subseteq G$. We say that for a $\oplus
$-orthogonal system $G=(x_{\kappa})_{\kappa\in H}$ the
element $\bigoplus G$ exists if\/f
$\bigvee\{\bigoplus K
\mid
K\subseteq G$ is f\/inite$\}$ exists in $E$ and then we put
$\bigoplus G=\bigvee\{\bigoplus K\mid K\subseteq G$ is
f\/inite$\}$. (Here we write $G_1\subseteq G$ if\/f there is
$H_1\subseteq H$ such that $G_1=(x_{\kappa})_{\kappa\in
H_1}$).

An element $u\in E$ is called {\em finite\/}
if either $u=0$ or there is a f\/inite sequence $\{a_1,a_2,\dots,a_n\}$ of not
necessarily dif\/ferent atoms of~$E$ such that $u=a_1\oplus
a_2\oplus \dots\oplus a_n$. Note that any atom of~$E$ is evidently f\/inite.

For an element $x$ of an ef\/fect algebra $E$ we write
$\ord(x)=\infty$ if $nx=x\oplus x\oplus\dots\oplus x$ ($n$-times)
exists for every positive integer $n$ and we write $\ord(x)=n_x$
if $n_x$ is the greatest positive integer such that $n_xx$
exists in $E$.  An ef\/fect algebra $E$ is {\em Archimedean} if
$\ord(x)<\infty$ for all $x\in E$.

\begin{definition}\label{subef}
Let $E$ be an  ef\/fect algebra.
Then $Q\subseteq E$ is called a {\em sub-effect algebra} of  $E$ if{\samepage
 \begin{enumerate}\itemsep=0pt
\item[$(i)$] $0, 1\in Q$,
\item[$(ii)$] if $x,y\in Q$ then $x'\in Q$ and
$x\bot y$ $\Longrightarrow$ $x\oplus y\in Q$.
\end{enumerate}}

\noindent
If $E$ is a lattice ef\/fect algebra and $Q$ is a sub-lattice and a sub-ef\/fect
algebra of $E$ then $Q$ is called a {\em sub-lattice effect algebra} of $E$.
\end{definition}

Note that a sub-ef\/fect algebra $Q$
(sub-lattice ef\/fect algebra $Q$) of an  ef\/fect algebra $E$
(of a~lattice ef\/fect algebra $E$) with inherited operation
$\oplus$ is an  ef\/fect algebra (lattice ef\/fect algebra)
in its own right.

Let $E$ be an ef\/fect algebra and let $(E_\kappa)_{\kappa\in
H}$ be a family of sub-ef\/fect algebras of $E$ such that:
\begin{description}\itemsep=0pt
\item[$(i)$]
$E=\bigcup\limits_{\kappa\in H} E_\kappa$.
\item[$(ii)$]
If $x\in E_{\kappa_1}\setminus\{0,1\}$,
$y\in E_{\kappa_2}\setminus\{0,1\}$ and
$\kappa_1\ne\kappa_2$, $\kappa_1,\kappa_2\in H$, then
$x\wedge y=0$ and $x\vee y=1$.
\end{description}
Then $E$ is called the  \emph{horizontal sum} of ef\/fect algebras
$(E_\kappa)_{\kappa\in H}$.

Important sub-lattice ef\/fect algebras of
a lattice ef\/fect algebra $E$ are
\begin{enumerate}\itemsep=0pt
\item[$(i)$] $S(E)=\{x\in E\mid x\wedge x'=0\}$  the
 {\em set of all sharp elements  of}  $E$ (see \cite{gudder1,gudder2}),
which is an orthomodular lattice (see~\cite{ZR57}).
\item[$(ii)$] Maximal subsets of pairwise compatible
elements of $E$ called {\em blocks} of $E$ (see \cite{ZR56}),
which are in fact maximal sub-$MV$-algebras of $E$.

\item[$(iii)$] The {\em center of compatibility} $B(E)$ of $E$,
 $B(E)=\bigcap\{M\subseteq E\mid M \
\mbox{is a block}$\ $\mbox{of}\ E\}=\{x\in E\mid x\comp y\
\mbox{for every}\ y\in E\}$ which is in fact
an $MV$-algebra ($MV$-ef\/fect algebra).
\item[$(iv)$]  The {\em center}
$C(E)=\{ x\in E \mid y=(y\wedge x)\vee(y\wedge x')\
\mbox{for all}\ y\in E\}$ of $E$ which is a~Boolean algebra
(see \cite{GrFoPu}). In every lattice ef\/fect algebra
it holds  $C(E)=B(E)\cap S(E)$ (see \cite{ZR51,ZR52}).
\end{enumerate}

For a poset $P$ and its subposet $Q\subseteq P$ we denote,
for all $X\subseteq Q$, by $\bigvee_{Q} X$ the join of
the subset $X$ in the poset $Q$ whenever it exists.

For a study  of ef\/fect algebras, we refer to \cite{dvurec}.

\section{Finite elements, modularity and atomicity\\ in lattice ef\/fect
algebras}

It is quite natural, for  a  lattice ef\/fect algebra,
to investigate whether the join of two f\/inite elements
is again f\/inite and whether each element below a f\/inite
element is again f\/inite. The following example shows that
generally it is not the case.

\begin{example}  Let $B$ be an inf\/inite complete atomic Boolean algebra,
$C$ a f\/inite chain $MV$-algebra. Then
\begin{enumerate}\itemsep=0pt
\item The set $F$ of f\/inite elements of the horizontal sum of $B$ and $C$
is not closed under order (namely, the top element $1$ is f\/inite but the coatoms
from $B$ are not f\/inite).
\item The set $F$ of f\/inite elements of the horizontal sum of two copies of~$B$
is not closed under join (namely, the join of two atoms in dif\/ferent copies of $B$ is the top element which is not f\/inite).
\end{enumerate}
\end{example}

\begin{theorem}\label{scompfinchain}  Let $E$ be a modular
lattice effect algebra, $x\in E$.
Then
\begin{enumerate}\itemsep=0pt  \item[$(i)$]
If $x$ is finite,  then, for every $y\in E$, $(x\vee y)\ominus y$ is finite {\rm \cite[Proposition 2.16]{ABVW}}.

 \item[$(ii)$] If $x$ is finite,  then every $z\in E$,
$z\leq x$ is finite.

 \item[$(iii)$] If $x$ is finite,  then every
 chain in the interval $[0, x]$ is finite.

 \item[$(iv)$] If $x$ is finite,
then  $[0, x]$ is a complete lattice.

 \item[$(v)$] If $x$ and $y$ are finite,  then  $x\vee y$ is finite.

 \item[$(vi)$]  The set $F$ of all finite elements of $E$
is  a lattice ideal of $E$.
\end{enumerate}
\end{theorem}
\begin{proof}  $(i)$  See \cite[Proposition 2.16]{ABVW}.

 $(ii)$ This follows at once from \cite[Proposition 2.16]{ABVW}
by putting $y=x\ominus z$.

 $(iii)$ This is an immediate consequence
of~\cite[Proposition 2.15]{ABVW} that gives a characterization
of f\/inite elements in modular lattice ef\/fect algebras using the height function
and of general and well-known facts about modular lattices (which can be
found, for instance, in \cite[\S~VII.4]{jacobson} and are also
recalled in \cite[\S~2.2, page~5]{ABVW}).

 $(iv)$ Since the interval $[0, x]$ has no inf\/inite chains it is complete
by \cite[Theorem 2.41]{davey}.

 $(v)$ Indeed $x \vee y =((x \vee y) \ominus y )\oplus y $
and, clearly, the sum of f\/inite elements is f\/inite.

 $(vi)$ It follows immediately from the above facts.
\end{proof}

Special types of ef\/fect algebras called sharply dominating and
$S$-dominating have been introduced by S.~Gudder in \cite{gudder1,gudder2}. Important example is the standard Hilbert spaces ef\/fect algebra~$\cal E(H)$.

\begin{definition}[\cite{gudder1,gudder2}]
An ef\/fect algebra $(E, \oplus, 0,
1)$ is called {\em sharply dominating} if for every $a\in E$ there
exists a smallest sharp element $\hat{a}$ such that $a\leq
\hat{a}$. That is $\hat{a}\in S(E)$ and if $b\in S(E)$ satisf\/ies
$a\leq b$ then $\hat{a}\leq b$.
\end{definition}

Similarly to \cite[Theorem 2.7]{PR3} we have the following.

\begin{theorem}\label{modlea}
Let $E$ be a modular  Archimedean lattice effect algebra
and let $E_1=\{x\in E \mid x\ \mbox{is} \ \mbox{finite}
\ \mbox{or}\ x' \   \mbox{is}  \ \mbox{finite}\}$.  Then
\begin{enumerate}\itemsep=0pt
\item[$(i)$]  $E_1$ is a sub-lattice effect algebra of $E$.
\item[$(ii)$] For every finite $x\in E$,
there exist a smallest
sharp element $\widehat{x}$ over $x$ and
a greatest sharp element $\widetilde{x}$ under $x$.
\item[$(iii)$] $E_1$ is sharply dominating.
\end{enumerate}
\end{theorem}
\begin{proof} $(i)$: Clearly, $x\in E_1$ if\/f $x'\in E_1$ by def\/inition of $E_1$.
Further for any f\/inite $x, y\in E_1$ we have  by
Theorem \ref{scompfinchain} that $x\vee y\in E_1$ and $x\oplus y\in E_1$ whenever
$x\oplus y$ exists. The rest follows by de Morgan laws and the fact that
$v\leq u$, $u$ is f\/inite implies $v$ is f\/inite (Theorem~\ref{scompfinchain}).

$(ii)$, $(iii)$: Let $x=\bigoplus_{i=1}^{n}k_i a_i$
for some set $\{a_1, \dots, a_n\}$ of atoms of $E$.
Clearly, for any index $j, 1\leq j\leq n$,
$k_j a_j\wedge \bigoplus_{i=1, i\not =j}^{n}k_i a_i=0$ and
$\bigoplus_{i=1, i\not =j}^{n}k_i a_i\leq (k_j a_j)'$.
Hence by \cite[Lemma 3.3]{PR5}
$n_{a_j} a_j\wedge \bigoplus_{i=1, i\not =j}^{n}k_i a_i=0$ and
$\bigoplus_{i=1, i\not =j}^{n}k_i a_i\leq (n_{a_j} a_j)'$.
By a successive application of the above argument this yields
the existence of the sum $\bigoplus_{i=1}^{n}n_{a_i} a_i$.
Then by
Theorem \ref{scompfinchain} the
interval $[0, \widehat{x}]$,
$\widehat{x}=\bigoplus_{i=1}^{n}n_{a_i} a_i$ is a complete
lattice ef\/fect algebra, hence it is sharply dominating.
Moreover by \cite[Theorem 3.5]{wujunde}, $\bigoplus_{i=1}^{n}n_{a_i} a_i$ is
the smallest sharp element $\widehat{x}$ over $x$.
It follows by  \cite{gudder1} that there exists
a greatest sharp element $\widetilde{x}$ under $x$ in
$[0, \widehat{x}]$ and so in $E$ and $E_1$ as well.

If $x'=\bigoplus_{i=1}^{m}l_i b_i$ for some set
$\{b_1, \dots, b_m\}$ of atoms of $E$ then
$w=\bigoplus_{i=1}^{m}n_{b_i} b_i$ is the smallest
sharp element  over $x'$. Hence $w'$
is the greatest sharp element under $x$ both in $E$ and $E_1$.
\end{proof}

Note that, in any ef\/fect algebra $E$, the following
inf\/inite distributive law holds (see \cite[Proposition 1.8.7]{dvurec}):
\[
\Big(\bigvee_{\alpha} c_{\alpha}\Big)\oplus b = \bigvee_{\alpha} (c_{\alpha}\oplus b)
\]
provided that $\bigvee_{\alpha} c_{\alpha}$ and  $(\bigvee_{\alpha} c_{\alpha})\oplus b$
exist.

 \begin{proposition}\label{xdist} Let $\{b_{\alpha} \mid \alpha\in \Lambda\}$
be a family of elements in a lattice effect algebra $E$ and let
$a\in E$ with $a\leq b_{\alpha}$ for all $\alpha\in \Lambda$. Then
\[
\Big(\bigvee\{b_{\alpha} \mid \alpha\in \Lambda\}\Big)\ominus a=%
\bigvee\{b_{\alpha}\ominus a \mid \alpha\in \Lambda\}
\]
if one side is defined.
\end{proposition}
\begin{proof}  Assume f\/irst that
$(\bigvee\{b_{\alpha} \mid \alpha\in \Lambda\})\ominus a$
is def\/ined. Then $\bigvee\{b_{\alpha} \mid \alpha\in \Lambda\}$ exists.
Clearly, $b_{\alpha} \ominus a\leq
(\bigvee\{b_{\alpha} \mid \alpha\in \Lambda\})\ominus a$ for all
$\alpha\in \Lambda$. Let $b_{\alpha} \ominus a\leq c$ for all
$\alpha\in \Lambda$. Let us put
$d=c\wedge ((\bigvee\{b_{\alpha} \mid \alpha\in \Lambda\})\ominus a)$.
Then $b_{\alpha} \ominus a\leq d$ for all
$\alpha\in \Lambda$ and
$d\leq (\bigvee\{b_{\alpha} \mid \alpha\in \Lambda\})\ominus a$. Hence
$d\oplus a$ exists and $b_{\alpha} \leq d\oplus a$
for all $\alpha\in \Lambda$. This yields
$\bigvee\{b_{\alpha} \mid \alpha\in \Lambda\}\leq d\oplus a$.
Consequently, $\bigvee\{b_{\alpha} \mid \alpha\in \Lambda\}\ominus a\leq d$,
so $d=\bigvee\{b_{\alpha} \mid \alpha\in \Lambda\}\ominus a\leq c$.

Now, assume that $\bigvee\{b_{\alpha}\ominus a \mid \alpha\in \Lambda\}$
is def\/ined.
Then $\bigvee\{b_{\alpha}\ominus a \mid \alpha\in \Lambda\}\leq 1\ominus a$,
which gives $\bigvee\{b_{\alpha}\ominus a \mid \alpha\in \Lambda\}\oplus a$ exists.
Hence by the above inf\/inite distributive law
\begin{gather*}
\bigvee\{b_{\alpha}\ominus a \mid \alpha\in \Lambda\}%
\oplus a=\bigvee\{(b_{\alpha}\ominus a) \oplus a\mid \alpha\in \Lambda\}%
=\bigvee\{b_{\alpha} \mid \alpha\in \Lambda\}.\tag*{\qed}
\end{gather*}\renewcommand{\qed}{}
\end{proof}

Now we are ready for the next proposition that was motivated by
\cite[\S~6, Theorem 20]{skornyakov} for complete modular lattices.

\begin{proposition}\label{xblockmod} Let $E$ be a modular lattice effect algebra,
$z\in E$
and let $F_z=\{ x \in E \mid x\ \mbox{is finite},\ x\leq z\}$
and suppose that $\bigvee F_z=z$.
Then the interval $[0, z]$ is atomic.
\end{proposition}

\begin{proof}  Let $0\not = y\in E$, $y\leq z$.
We shall show that there exists
an atom $a \leq y$. We have  (by the same argument as
in {\rm{}\cite[Lemma 3.1 (a)]{ABVW}} for complete lattices) that
\begin{gather*}
\bigvee\{(x\vee (z\ominus y))\ominus (z\ominus y)\mid x\in F_z\}\\
\qquad{}= \bigvee\{x\vee (z\ominus y)\mid x\in F_z\}\ominus (z\ominus y)=
z\ominus (z\ominus y)=y.
\end{gather*}
This yields $(x \vee (z\ominus y)) \ominus (z\ominus y) \not= 0$ for some $x \in F_z$.
By Theorem~\ref{scompfinchain}~$(i)$  $(x \vee (z\ominus y)) \ominus (z\ominus y)  \in F_z$,
$(x \vee (z\ominus y)) \ominus (z\ominus y)\leq y$. Hence,
there exists an atom $a \leq  (x \vee (z\ominus y)) \ominus (z\ominus y) \leq y$.
\end{proof}

\begin{corollary}\label{blockmod} Let $E$ be a modular lattice effect algebra
and let $F=\{ x \in E \mid x\ \mbox{is finite}\}$
and suppose that $\bigvee F=1$.
Then $E$ is atomic.
\end{corollary}

\begin{corollary}\label{dusblockmod} Let $E$ be a modular lattice effect algebra.
Let at least one block $M$
of $E$ be Archi\-me\-dean and atomic.
Then $E$ is atomic.
\end{corollary}
\begin{proof} Let us put $F=\{ x \in E \mid x\ \mbox{is f\/inite}\}$.
Clearly, $F$ contains all f\/inite elements of the block~$M$. Hence by
\cite[Theorem 3.3]{ZR65} we have $1=\bigvee_M (F\cap M)$.
From \cite[Lemma 2.7]{PR6} we obtain that the joins
in $E$ and $M$ coincide. Therefore $1=\bigvee_M (F\cap M)=%
\bigvee_E (F\cap M)\leq \bigvee_E F$. By Corollary
\ref{blockmod} we get that $E$ is atomic.
\end{proof}

Further recall that an element $u$ of a lattice $L$ is called
a {\em compact element} if, for any $D\subseteq L$ with $\bigvee D\in L$,
$u\leq \bigvee D$ implies  $u\leq \bigvee F$ for some f\/inite $F\subseteq D$.

Moreover, the lattice $L$ is called {\em compactly generated} if every
element of $L$ is a join of compact elements.

It was proved in \cite[Theorem 6]{PR4} that every
compactly generated lattice ef\/fect algebra is atomic.
If moreover $E$ is Archimedean then every compact
element $u\in E$ is f\/inite \cite[Lemma~4]{PR4} and conversely
\cite[Lemma~2.5]{PR6}.

\begin{example}[\protect{\cite[Example 2.9]{PR3}}] If $a$ is an atom of a
compactly generated Archimedean lattice ef\/fect algebra $E$ (hence atomic)
then $n_a a$ need not be an atom of $S(E)$.

Indeed, let $E$ be a horizontal sum of a Boolean
algebra $B=\{0, a, a', 1=a\oplus a'\}$
and a chain $M=\{0, b, 1=2b\}$. Then $S(E)=B$ and $1=2b$ is
not an atom of $S(E)$.
\end{example}

\begin{remark}
The atomicity of the set of sharp elements $S(E)$ is not completely
solved till now. For example, if $E$ is a complete modular Archimedean atomic lattice ef\/fect
algebra then $S(E)$ is an atomic orthomodular lattice (see \cite{PR6}).
\end{remark}

 This remark leads us to

\begin{proposition}\label{mose}
Let $E$ be a modular Archimedean atomic lattice effect
algebra. Then $S(E)$ is an atomic orthomodular lattice.
\end{proposition}
\begin{proof} Let $x\in S(E)$, $x\not= 0$. From
\cite[Theorem 3.3]{ZR65} we
get that there is an atom~$a$ of~$E$ such that
$n_a{}a \leq x$. Then by Theorem \ref{scompfinchain}
the interval $[0, n_a a]$ is a complete
modular atomic lattice ef\/fect algebra. This
yields that $[0, n_a a]$ is a compactly generated complete modular
lattice ef\/fect algebra and all elements of $[0, n_a a]$ are
compact in $[0, n_a a]$.
Hence also   $S([0, n_a a])$ is a compactly  generated complete modular
lattice ef\/fect algebra i.e.\ it is atomic. Clearly, any atom~$p$ of
$S([0, n_a a])$ is an atom of $S(E)$ and $p\leq x$.
\end{proof}

A {\em basic algebra} \cite{CHK1} ({\em lattice with sectional antitone involutions})
is a system
${L}=(L;\vee,\wedge,$ $(^a)_{a\in L},0,1)$, where
$(L;\vee,\wedge,0,1)$ is a bounded lattice such that every
principal order-f\/ilter $[a,1]$ (which is called a {\em section})
possesses an antitone involution $x\mapsto x^a$.

Clearly, any principal ideal 
$[0,x]$, $x\in L$ of a basic algebra
$L$  is again a basic algebra. Moreover, any lattice ef\/fect algebra is a
basic algebra.  Note that every interval $[a, b]$,
for $a < b$ in an ef\/fect algebra $E$ can be organized (in a natural way)
into an ef\/fect algebra (see \cite[Theorem 1]{sykes}), hence every interval
$[a, b]$ in $E$ possesses an antitone involution.


The following Lemma is in fact implicitly contained in the proof
 of \cite[Theorem 5]{PR4}.

\begin{lemma}[\protect{\cite[Theorem 5]{PR4}}]\label{crae} Let $L$ be a basic algebra, $u\in L$,
$u\not=0$ a compact element. Then there is an atom $a\in L$ such that
$a\leq u$.
\end{lemma}

\begin{lemma}\label{cisfin} Let $E$ be an Archimedean lattice
effect algebra, $u\in E$ a compact element. Then
$u$ is a finite join of finite elements of $E$.
\end{lemma}

\begin{proof} If $u=0$ we are f\/inished. Let $u\not=0$.
Let ${\mathcal Q}$ be a maximal pairwise compatible subset of f\/inite elements
of $E$ under $u$. Clearly, $0\in {\mathcal Q}\not=\varnothing$.
Assume that $u$ is not the smallest upper bound of ${\mathcal Q}$
in $E$. Hence  there is an
element $c\in E$ such that $c$ is an upper bound
of ${\mathcal Q}$, $c\not\geq u$. Let
us put $d=c\wedge u$. Then $d < u$. Clearly, the interval $[d, 1]$ is a basic algebra
and $u$ is compact in $[d, 1]$. Hence
by Lemma~\ref{crae}  there is an atom $b\in [d, 1]$  such that $b\leq u$.
Let us put $a=b\ominus d$. Then  $a$ is an atom of $E$. Let $M$  be a block of $E$
containing the compatible set ${\mathcal Q}\cup \{d, b, u\}$.
Evidently $a\in M$ and
${\mathcal Q}\cup \{q\oplus a \mid q\in {\mathcal Q}\}\subseteq M$ is a compatible
subset of f\/inite elements of $E$ under $u$. From the maximality of $\mathcal Q$
we get that $\{q\oplus a \mid q\in {\mathcal Q}\}\subseteq {\mathcal Q}$. Hence,
for all $n\in {\mathbb N}$, $na\in {\mathcal Q}$, a contradiction with the assuption
that $E$ is Archimedean. Therefore $u=\bigvee {\mathcal Q}$. Since~$u$ is compact
there are f\/initely many f\/inite elements $q_1, \dots, q_n$ of ${\mathcal Q}$
such that $u=\bigvee_{i=1}^{n} q_i$.
\end{proof}

\begin{remark}{\rm The condition that $E$  is Archimedean in Lemma \ref{cisfin}
cannot be omitted (e.g., the Chang $MV$-ef\/fect algebra
$E=\{0, a, 2a, 3a, \dots, (3a)', (2a)',a', 1\}$ is not Archimedean,
every  $x\in E$ is compact and the top element $1$
is not a f\/inite join of f\/inite elements of $E$). }
\end{remark}

\begin{corollary}\label{modcompis}
Let $E$ be a modular Archimedean lattice
effect algebra, $u\in E$ a compact element. Then
$u$ is finite.
\end{corollary}
\begin{proof}
Since $u$ is a f\/inite join of f\/inite elements of $E$ and in a modular
lattice ef\/fect algebra a~f\/inite join of f\/inite elements is f\/inite
by Theorem~\ref{scompfinchain} we are done.
\end{proof}

Thus we obtain the following common corollary of
Proposition~\ref{xblockmod} and Corollary~\ref{modcompis}.

\begin{theorem}\label{compblockmod} Let $E$ be a modular Archimedean lattice effect algebra,
$z\in E$
and let $C_z=\{ x \in E \mid x\ \mbox{is compact}, x\leq z\}$
and suppose that $\bigvee C_z=z$.
Then the interval $[0, z]$ is atomic. Moreover, if $z=1$ and $\bigvee C_1=1$
then $E$ is atomic.
\end{theorem}

\section{States on modular Archimedean atomic lattice ef\/fect
algebras}

The aim of this section is to apply results of previous section
in order to study $(o)$-continuous states on
modular Archimedean  atomic  lattice ef\/fect algebras.

\begin{definition}\label{Dstate}
Let $E$ be an ef\/fect algebra. A map $\omega:E\to[0,1]$ is called a
{\em state} on $E$ if $\omega(0)=0$, $\omega(1)=1$ and $\omega(x\oplus
y)=\omega(x)+\omega(y)$ whenever $x\oplus y$ exists in $E$.
If, moreover, $E$ is lattice ordered then $\omega$  is
called {\em subadditive} if $\omega(x\vee y)\leq \omega(x)+\omega(y)$,
for all $x, y\in E$.
\end{definition}

It is easy to check that the notion of a state $\omega$ on an orthomodular
lattice $L$ coincides with the notion of a state on its derived ef\/fect algebra
$L$. It is because $x\le y'$ if\/f $x\oplus y$ exists in $L$, hence $\omega(x\vee
y)=\omega(x\oplus y)=\omega(x)+\omega(y)$ whenever $x\le y'$ (see \cite{kalmbach}).

It is easy to verify that, if $\omega$ is a subadditive state
on a lattice ef\/fect algebra E, then in fact
$\omega(x)+\omega(y) = \omega(x\vee y)+\omega(x \wedge y)$
for all $x, y \in E$ (see \cite[Theorem 2.5]{ZR63}),
so that $\omega$ is a {\em modular measure}, as def\/ined
for example in \cite[\S~5, page 13]{ABVW}.

Assume that $(\mathcal E;\prec)$ is a directed set and $E$
is an ef\/fect algebra. A net of elements of $E$ is denoted by
$(x_{\alpha})_{\alpha\in\mathcal E}$.
Then $x_{\alpha}\uparrow  x$ means that
$x_{\alpha_1}\leq x_{\alpha_2}$ for every
${\alpha_1}\prec {\alpha_2}$, ${\alpha_1}, {\alpha_2}\in {\mathcal E}$
and $x=\bigvee\{ x_{\alpha}\mid {\alpha}\in {\mathcal E}\}$.
The meaning of  $x_{\alpha}\downarrow x$ is dual.
A  net $(x_{\alpha})_{\alpha\in\mathcal E}$
of elements of an ef\/fect algebra
$E$  {\em order converges to a point
$x\in E$} if there are nets $(u_{\alpha})_{\alpha\in\mathcal E}$ and
$(v_{\alpha})_{\alpha\in\mathcal E}$ of elements of~$E$ such that
\[
u_{\alpha}\uparrow x,\ v_{\alpha}\downarrow x, \
\mbox{and}\ u_{\alpha}\le x_{\alpha}\le v_{\alpha}\
\mbox{for all}\ {\alpha\in\mathcal E}.
\]
We write {\em $x_{\alpha}\oo x$, ${\alpha\in\mathcal E}$ in $E$}
(or brief\/ly $x_{\alpha}\oo x$).

A state $\omega$ is called
{\em$(o)$-continuous} ({\em order-continuous\/})  if, for every net
$(x_{\alpha})_{\alpha\in \mathcal{E}}$ of elements of~$E$,
$x_{\alpha}\too x$  $\implies$
$\omega(x_{\alpha})\to \omega(x)$ (equivalently
$x_\alpha\uparrow x\Rightarrow
\omega(x_\alpha)\uparrow \omega(x)$).

We are going to prove statements about the existence of
$(o)$-continuous states which are subadditive.

\begin{theorem}\label{exstatecen}
Let $E$ be a Archime\-dean atomic lattice
effect algebra, $c \in C(E)$, $c$ finite in $E$, $c\not =0$, $[0, c]$ a modular lattice.
Then there exists an $(o)$-continuous state $\omega$ on $E$,
which is subadditive.
\end{theorem}
\begin{proof}  Note that, for every central element
$z$ of a lattice ef\/fect algebra $E$,
the interval $[0,z]$ with the $\oplus$ operation inherited from
$E$ and the new unit $z$ is a lattice ef\/fect algebra in its own
right.

Since $c$ is central we have
have the
direct product decomposition $E\cong [0, c]\times [0, c']$. Hence
$E=\{y\oplus z \mid y\in [0, c], z\in [0, c']\}$.

Since $c$ is f\/inite in $E$ and hence in $[0, c]$
we have by Theorem \ref{scompfinchain}
that the interval $[0, c]$ is a~complete modular atomic lattice
ef\/fect algebra. From \cite[Theorem 4.2]{ZR74} we get a subadditive
$(o)$-continuous state $\omega_c$ on $[0, c]$.

Let us def\/ine
$\omega:E\to [0, 1]\subseteq {\mathbb{R}}$
by setting $\omega(x)=\omega_c(y)$,
for every $x=y\oplus z$, $y\in [0, c], z\in [0, c']$.
It is easy to check that $\omega$ is
an $(o)$-continuous state  on $E$, which is subadditive.
These properties follow by the fact that the ef\/fect algebra operations
as well as the lattice operations on the direct product
$[0, c]\times [0, c']$ are def\/ined coordinatewise and
$\omega_c$ is a state on the complete modular atomic lattice ef\/fect algebra
$[0, c]$ with all enumerated properties. \end{proof}

\begin{corollary}\label{corexstatecen}
Let $E$ be a modular Archime\-dean atomic lattice
effect algebra, $c \in C(E)$, $c$ finite in $E$, $c\not =0$.
Then there exists an $(o)$-continuous state $\omega$ on $E$,
which is subadditive.
\end{corollary}

In \cite{ZR57} it was proved
that for every lattice ef\/fect algebra $E$ the subset $S(E)$
is an orthomodular lattice. It follows that $E$ is an orthomodular lattice
if\/f $E=S(E)$. If $E$ is atomic then $E$ is an orthomodular lattice
if\/f $a\in S(E)$ for every atom $a$ of $E$. This is because if
$x\in E$ with $x\wedge x'\not =0$ exists then there exists an atom $a$
of $E$ with $a \leq x\wedge x'$, which gives $a\leq x' \leq a'$ and hence
$a\wedge a'=a\not =0$, a contradiction.

\begin{theorem}\label{exstate}
Let $E$ be an Archime\-dean atomic lattice
effect algebra with $S(E)\not = E$. Let
$F=\{ x \in E \mid x\ \mbox{is finite}\}$ be an ideal of $E$ such that
$F$ is a modular lattice.
Then there exists an $(o)$-continuous state $\omega$ on $E$,
which is subadditive.
\end{theorem}
\begin{proof} Let $x\in E\setminus S(E)$. From Theorem \cite[Theorem 3.3]{ZR65}
we have that there are mutually distinct atoms $a_{\alpha}\in
E$ and positive integers $k_{\alpha}$, $\alpha\in{\cal E}$ such
that
$
x=\bigoplus\{k_{\alpha}a_{\alpha}\mid
\alpha\in{\cal E}\}=
\bigvee\{k_{\alpha}a_{\alpha}\mid
\alpha\in{\cal E}\},
$
and $x\in S(E)$ if\/f
$k_{\alpha}=n_{a_{\alpha}}=\ord(a_{\alpha})$ for all
$\alpha\in\cal E$. Hence there is an atom $a\in E$ such that
$a\not\in S(E)$ i.e., $a\le a'$.

We shall proceed similarly as in \cite[Theorem 3.1]{ZR74}.

 $(i)$: Assume that $a\in B(E)$. Then also $n_a a\in B(E)$ (by \cite{ZR56}) and, by
\cite[Theorem~2.4]{ZR70}, $n_aa\in S(E)$. Thus $n_aa\in B(E)\cap
S(E)=C(E)$.

$(ii)$: Assume  now that $a\not\in B(E)$. Then
there exists an atom $b\in E$
with $b\notcomp a$. As $F$ is a~modular lattice  we have
$[0,b]=[a\wedge b,b]\cong[a,a\vee b]$ which yields that $a\vee b$ covers $a$
both in $F$ and $E$.
Hence there exists an atom $c\in E$ such that
$a\oplus c=a\vee b$, which gives $c\le a'$.
Evidently, $c\ne b$ as $b\not\le a'$. If
$c\ne a$ then $a\vee c = a\oplus c=a\vee b$, which implies $b\le a\vee c\le
a'$, a contradiction. Thus $c=a$ and $a\vee b=2a$.

Let $p\in E$ be an atom. Then either $p\notcomp a$ which, as we
have just shown, implies that $p\le p\vee a=2a\le n_aa$, or
$p\comp a$ and hence $p\comp n_a a$ for every atom $p\in E$. By
\cite{ZR60}, for every $x\in E$ we have $x=\bigvee\{u\in E\vert
u\le x$, $u$ is a sum of f\/inite sequence of atoms$\}$. Since
$n_a a\comp p$ for every atom $p$ and hence $n_a a\comp u$ for every
f\/inite sum $u$ of atoms, we conclude  that
$n_a a\comp x$ for every $x\in E$. Thus again $n_aa\in B(E)\cap
S(E)=C(E)$.

Since $n_a a$ is f\/inite and the interval $[0, n_a a]$ is modular
we can apply  Theorem \ref{exstatecen}.
\end{proof}

\begin{corollary}\label{xexstate}
Let $E$ be a modular Archime\-dean atomic lattice
effect algebra with $S(E)\not = E$.
Then there exists an $(o)$-continuous state $\omega$ on $E$,
which is subadditive.
\end{corollary}
\begin{proof}
It follows immediately from Theorem \ref{scompfinchain}~$(vi)$
and Theorem~\ref{exstate}.
\end{proof}

\subsection*{Acknowledgements}

We gratefully
acknowledge f\/inancial support
of the  Ministry of Education of the Czech Republic
under the project MSM0021622409.
We also thank the anonymous referees for the very thorough
reading and contributions to improve
our presentation of the paper.

\pdfbookmark[1]{References}{ref}
\LastPageEnding


\begin{thebibliography}{99}

\footnotesize\itemsep=0pt

\bibitem{ABVW}
Avallone A., Barbieri G., Vitolo P., Weber H.,
Decomposition of ef\/fect algebras and the Hammer--Sobczyk theorem,
\href{http://dx.doi.org/10.1007/s00012-008-2083-z}{{\it Algebra Universalis}} {\bf 60} (2009), 1--18.

\bibitem{BGL}
Busch P., Grabowski M., Lahti P.J.,
Operational quantum physics, {\it Lecture Notes in Physics. New Series~m: Monographs}, Vol.~31, Springer-Verlag, New York, 1995.


\bibitem{CHK1} Chajda I., Hala{\v{s}} R., K\"u{}hr J.,
Many-valued quantum algebras,
\href{http://dx.doi.org/10.1007/s00012-008-2086-9}{{\it Algebra Universalis}} {\bf 60}  (2009), 63--90.


\bibitem{C.C.Ch}
Chang C.C.,
Algebraic analysis of many valued logics,
\href{http://dx.doi.org/10.2307/1993227}{{\it Trans. Amer. Math. Soc.}} {\bf 88} (1958), 467--490.

\bibitem{Kop3}
 Chovanec F., K\^opka F.,
Dif\/ference posets in the quantum structures background,
\href{http://dx.doi.org/10.1023/A:1003625401906}{{\it Internat. J. Theoret. Phys.}} {\bf 39} (2000), 571--583.

 \bibitem{davey}
 Davey B.A., Priestley H.A.,
Introduction to lattices and order, 2nd ed.,
Cambridge University Press, New York, 2002.

\bibitem{dvurec}
Dvure{\v{c}}enskij A., Pulmannov\'{a} S.,
New trends in quantum structures, {\it Mathematics and its Applications}, Vol.~516, Kluwer Academic Publishers, Dordrecht; Ister Science, Bratislava, 2000.

\bibitem{dvurec2}
Dvure{\v{c}}enskij A., Graziano M.G.,
An invitation to economical test spaces and ef\/fect algebras,
\href{http://dx.doi.org/10.1007/s00500-004-0365-8}{{\it Soft Comput.}} {\bf 9} (2005), 463--470.

\bibitem{EpZh}
Epstein L.G., Zhang J.,
Subjective probabilities on subjectively unambiguous events,
\href{http://dx.doi.org/10.1111/1468-0262.00193}{{\it Econometrica}} {\bf 69} (2001), 265--306.

\bibitem{FoBe}
Foulis D.J., Bennett M.K., Ef\/fect algebras and unsharp quantum logics,
\href{http://dx.doi.org/10.1007/BF02283036}{{\it Found. Phys.}} {\bf 24} (1994), 1331--1352.

\bibitem{foulis2007}
Foulis D.J.,
Ef\/fects, observables, states, and symmetries in physics,
\href{http://dx.doi.org/10.1007/s10701-007-9170-4}{{\it Found. Phys.}} {\bf 37} (2007),  1421--1446.


\bibitem{GrFoPu}
Greechie R.J., Foulis D.J., Pulmannov\'a S.,
The center of an ef\/fect algebra,
\href{http://dx.doi.org/10.1007/BF01108592}{{\it Order}} {\bf 12} (1995), 91--106.

\bibitem{gudder1}
 Gudder S.P.,
 Sharply dominating ef\/fect algebras,
{\it Tatra Mt. Math. Publ.} {\bf 15} (1998), 23--30.

\bibitem{gudder2}
Gudder S.P.,
$S$-dominating ef\/fect algebras,
\href{http://dx.doi.org/10.1023/A:1026637001130}{{\it Internat. J. Theoret. Phys.}} {\bf 37} (1998), 915--923.

 \bibitem{jacobson}
Jacobson N.,
Lectures in abstract algebra, Vol.~I, Basic concepts, D. Van Nostrand Co., Inc., Toronto~-- New York~-- London, 1951.


\bibitem{ZR57}
Jen\v{c}a G., Rie\v{c}anov\'a Z.,
On sharp elements in lattice ordered ef\/fect algebras,
{\it BUSEFAL} {\bf 80}  (1999), 24--29.

\bibitem{kalmbach}
Kalmbach G.,
Orthomodular lattices, {\it Mathematics and its Applications}, Vol.~453, Kluwer Academic Publishers, Dordrecht, 1998.


\bibitem{Kop}
K\^opka F., D-posets of fuzzy sets,
{\it Tatra Mt. Math. Publ.} {\bf 1} (1992), 83--87.

\bibitem{KoCh}
K\^opka F., Chovanec F.,
$D$-posets,
{\it Math. Slovaca} {\bf 44}  (1994), 21--34.

\bibitem{Kop2}
K\^opka F.,
Compatibility  in $D$-posets,
\href{http://dx.doi.org/10.1007/BF00676263}{{\it Internat. J. Theoret. Phys.}} {\bf 34} (1995), 1525--1531.

\bibitem{PR4}
Paseka J., Rie\v canov\'a Z.,
Compactly generated de Morgan lattices, basic algebras and ef\/fect algebras,
\href{http://dx.doi.org/10.1007/s10773-009-0011-4}{{\it Internat. J. Theoret. Phys.}}, to appear.

 \bibitem{PR5}
 Paseka J., Rie\v canov\'a Z., Wu  J.,
 Almost orthogonality and Hausdorf\/f interval topologies of atomic lattice ef\/fect algebras,  \href{http://arxiv.org/abs/0908.3288}{arXiv:0908.3288}.

\bibitem{PR6}
Paseka J., Rie\v canov\'a Z.,
The inheritance of BDE-property in sharply dominating  lattice ef\/fect algebras and $(o)$-{continuous} states,
{\it Soft Comput.}, to appear.



\bibitem{ZR51}
Rie\v{c}anov\'{a} Z.,
Compatibility and central elements in ef\/fect algebras,
{\it Tatra Mt. Math. Publ.} {\bf 16} (1999), 151--158.

\bibitem{ZR52}
Rie\v{c}anov\'{a} Z.,
Subalgebras, intervals and central elements of generalized ef\/fect algebras,
\href{http://dx.doi.org/10.1023/A:1026682215765}{{\it Internat. J. Theoret. Phys.}} {\bf 38} (1999), 3209--3220.

\bibitem{ZR56}
Rie\v{c}anov\'{a} Z.,
Generalization of blocks for $D$-lattices and lattice-ordered ef\/fect algebras,
\href{http://dx.doi.org/10.1023/A:1003619806024}{{\it Internat. J. Theoret. Phys.}} {\bf 39} (2000),  231--237.

\bibitem{ZR60}
Rie\v{c}anov\'{a} Z.,
Orthogonal sets in ef\/fect algebras,
{\it Demonstratio Math.} {\bf 34} (2001),  525--532.


\bibitem{ZR62}
Rie\v{c}anov\'{a} Z.,
Proper ef\/fect algebras admitting no states,
\href{http://dx.doi.org/10.1023/A:1011911512416}{{\it Internat. J. Theoret. Phys.}} {\bf 40} (2001), 1683--1691.

 \bibitem{ZR63}
 Rie\v{c}anov\'{a} Z.,
Lattice ef\/fect algebras with $(o)$-continuous  faithful valuations,
\href{http://dx.doi.org/10.1016/S0165-0114(01)00102-6}{{\it Fuzzy Sets and Systems}}  {\bf 124} (2001), no.~3, 321--327.

\bibitem{ZR65}
Rie\v{c}anov\'{a} Z.,
Smearings of states def\/ined on sharp elements onto ef\/fect algebras,
\href{http://dx.doi.org/10.1023/A:1020136531601}{{\it Internat. J. Theoret. Phys.}} {\bf 41}  (2002), 1511--1524.

\bibitem{ZR70}
Rie\v{c}anov\'a Z.,
Continuous lattice ef\/fect algebras admitting order-continuous states,
\href{http://dx.doi.org/10.1016/S0165-0114(02)00141-0}{{\it Fuzzy Sets and Systems}} {\bf 136} (2003), 41--54.

\bibitem{ZR74}
Rie\v{c}anov\'a Z.,
Modular atomic ef\/fect algebras and the existence of subadditive states,
{\it Kybernetika} {\bf 40} (2004), 459--468.


\bibitem{PR3}
Rie\v{c}anov\'a Z., Paseka J.,
State smearing theorems and the existence of states on some atomic lattice ef\/fect algebras,
\href{http://dx.doi.org/10.1093/logcom/exp018}{{\it J. Logic Comput.}}, to appear.

\bibitem{wujunde}
Rie\v{c}anov\'{a} Z., Wu  J.,
States on sharply dominating ef\/fect algebras,
\href{http://dx.doi.org/10.1007/s11425-007-0163-8}{{\it Sci. China Ser. A}} {\bf 51} (2008), 907--914.


\bibitem{skornyakov}
Skornyakov L.A.,
Elements of lattice theory, 2nd ed., Nauka, Moscow, 1982
(English transl.: 1st ed., Hindustan, Delhi; Adam Hilger, Bristol, 1977).

\bibitem{sykes}
Sykes S.R.,
Finite modular ef\/fect algebras,
\href{http://dx.doi.org/10.1006/aama.1997.0543}{{\it Adv. in Appl. Math.}} {\bf 19} (1997), 240--250.

\end{thebibliography}
\end{document}